\documentclass[twocolumn,floats,showpacs,amsmath,amssymb,prb]{revtex4}
\usepackage{graphicx}
\begin{document}

\title{Isotope effects on the lattice parameter of cubic SiC}
\author{Carlos P. Herrero}
\author{Rafael Ram\'{\i}rez}
\affiliation{Instituto de Ciencia de Materiales de Madrid,
         Consejo Superior de Investigaciones Cient\'{\i}ficas (CSIC),
         Campus de Cantoblanco, 28049 Madrid, Spain }
\author{Manuel Cardona}
\affiliation{Max-Planck-Institut f\"ur Festk\"orperforschung,
         Heisenbergstrasse 1, 70569 Stuttgart, Germany}
\date{\today}

\begin{abstract}
Path-integral molecular dynamics simulations in the isothermal-isobaric 
($NPT$) ensemble have been carried out to study the dependence of the lattice
parameter of 3C-SiC upon isotope mass.
This computational method allows a quantitative and nonperturbative study 
of such anharmonic effect.
Atomic nuclei were treated as quantum particles
interacting via a tight-binding-type potential.
At 300 K, the difference $\Delta a$ between lattice parameters of
3C-SiC crystals with $^{12}$C and $^{13}$C amounts to 
$2.1 \times 10^{-4}$~\AA. The effect due to Si isotopes is
smaller, and amounts to $3.5 \times 10^{-5}$~\AA \ when replacing 
$^{28}$Si by $^{29}$Si.
Results of the PIMD simulations are interpreted in terms of
a quasiharmonic approximation for the lattice vibrations.
\end{abstract}

\pacs{71.20.Nr, 71.15.Pd, 63.20.Ry, 63.20.Dj}


\maketitle

 It is well known that
 the lattice parameters of two chemically identical crystals with different
isotopic composition are not equal, lighter isotopes giving rise to
larger lattice parameters. This is caused by a combination of two factors:
the dependence of atomic vibrational amplitudes upon atomic mass, and the 
anharmonicity of the vibrations.
The isotope effect on the lattice parameter is most important at low temperatures, 
due to the change of zero-point vibrational amplitude with
atomic mass, and disappears in the high-temperature
(classical) limit at $T > \Theta_D$ ($\Theta_D$, Debye temperature),
where vibrational amplitudes are independent of the mass.
In recent years, it has become feasible to measure the isotopic effect in 
lattice parameters of crystals with high precision.\cite{ka98}
Most of the work has been performed on elemental crystals, although binary
and multinary materials offer the attractive possibility of isotopic 
substitution on different atoms.\cite{de96}
Of these materials, SiC, with 3 stable isotopes of Si and 2 of C,
plus about 70 different polytypes, is a paragon. Here, however, we confine
ourselves to the simplest polytype: zincblende-like 3C-SiC.

SiC has been suggested for a number of applications exploiting many of its
superlative properties, close to those of diamond.\cite{ch03b,sa05}
Some of these applications take advantage of its hardness, large thermal
conductivity, and low thermal expansion. 
Isotopically modified SiC may find applications exploiting the higher thermal
conductivity\cite{be93} and the dependence of its hardness on isotopic 
composition.\cite{ra93}

 Anharmonic effects in the vibrational properties of 3C-SiC have
been studied earlier in detail, e.g.,
the pressure and temperature dependence of phonon
frequencies\cite{ol82a,ol82b} and lifetimes.\cite{de01,ka96}
The thermal expansion of cubic SiC (another anharmonic effect) has
been studied in detail both experimentally\cite{sl75} and
theoretically\cite{tal95,ka94}, the latter using a
quasiharmonic approximation (QHA) for the lattice vibrations.

Isotopic effects on the lattice parameters of crystals have been
usually calculated by employing the QHA and perturbative 
methods based on {\em ab-initio} techniques.\cite{pa94b,bi94,de96}
An alternative to perturbational approaches
in solids is the combination of the path integral formulation
(to deal with the quantum nature of the nuclei)
with electronic structure methods.
The path integral approach to statistical mechanics
allows one to study finite-temperature
properties of quantum many-body problems in a nonperturbative
scheme, even in the presence of large anharmonicities.\cite{ce95}
An advantage of its combination with electronic structure methods
is that both electrons and atomic
nuclei are treated quantum mechanically in the framework of the
Born-Oppenheimer approximation.\cite{tu97,ra98,ch03}
The path-integral molecular dynamics (PIMD)
 method is based on an isomorphism between the quantum
system under consideration and a classical one, obtained by replacing
each quantum particle by a cyclic chain of classical particles,
connected by harmonic springs.\cite{gi88,ce95,no96}

When calculating properties of crystals with isotopically mixed
composition, it is usually assumed that each atomic nuclei in the solid
has a mass equal to the average mass. This kind of {\em virtual-crystal
approximation} has been used in density-functional calculations, as well
as in atomistic simulations based on path 
integrals.\cite{de96,ca05b,no96,he99,he01b}
In fact, in earlier path integral simulations of diamond it was found
that the results obtained by using this approximation are indistinguishable
from those derived from simulations in which actual isotopic mixtures
were considered.\cite{he01b} 

Here we extend earlier path integral calculations of the lattice parameter 
of group-IV solids\cite{no96,he99,he01b} (diamond, Si, Ge) 
to a IV-IV compound such as 3C-SiC. 
The electronic structure has been treated with an efficient
tight-binding Hamiltonian, based on density-functional 
calculations.\cite{po95,ra08}
Simulations were performed on a $2\times2\times2$ supercell of the
3C-SiC face-centered cubic cell with periodic boundary conditions,
including 64 atoms.  For a given temperature and isotopic composition, 
a typical run consisted of $2 \times 10^4$ PIMD 
steps for system equilibration, followed by $4 \times 10^6$ steps for the 
calculation of ensemble average properties.
Details on the actual implementation of the PIMD method to study
structural and electronic properties of SiC were given
elsewhere.\cite{ra08}

First of all, we quantify the influence of quantum effects on the
lattice parameter of cubic SiC. With this purpose we have performed 
simulations in which the atomic nuclei of either C or Si (or both) 
were considered as classical particles.
We note that in the formalism used here, the classical limit for a 
given atomic nucleus is obtained by taking its mass as $M \to \infty$
(in fact, we took masses in the order of $10^5$ amu).

\begin{figure}
\vspace{-2.0cm}
\hspace{-0.5cm}
\includegraphics[width= 9cm]{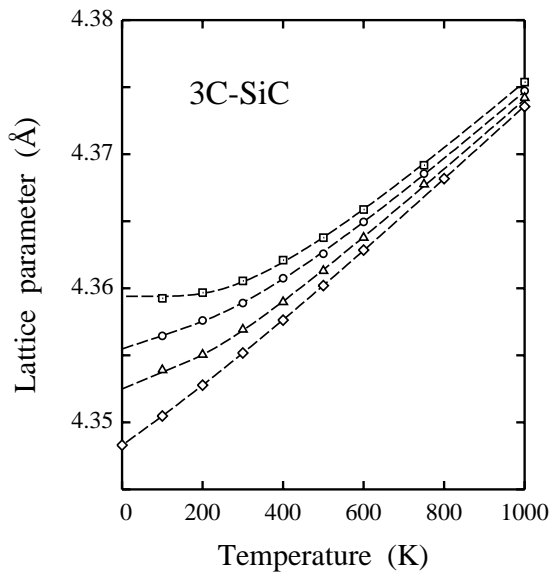}
\vspace{-2.5cm}
\caption{
 Temperature dependence of the lattice parameter of 3C-SiC, as obtained
from (a)~PIMD simulations (squares), (b) simulations
performed by setting the Si nuclei as classical particles (circles),
(c) simulations carried out by considering classical C nuclei (triangles),
and (d) all nuclei are assumed to be classical (diamonds).
Lines represent empirical fits.
The statistical error of the simulation results is less than the symbol
size.
}
\label{fig1}
\end{figure}

In Fig.1 we present the results of our molecular dynamics simulations
for the various cases considered. First, we compare the results of
the full PIMD simulations (quantum atomic nuclei, open squares) with those 
of classical simulations (diamonds). The zero-point motion induces an
increase in the lattice parameter of 0.011 \AA, which means a relative
change of $2.5 \times 10^{-3}$. Second, we are interested in the lattice
parameter obtained when nuclei of one of the elements (Si or C) are 
considered as classical and the other as quantum particles. 
Thus, triangles represent results for
quantum Si and classical C, whereas circles indicate data for quantum C
and classical Si. 
By comparing these results with those found in the full PIMD simulations,
we find for $T \to 0$ that the lattice parameter $a$ increases by 
$3.9 \times 10^{-3}$~\AA\ and 
$6.8 \times 10^{-3}$~\AA, due to having considered Si or C as quantum 
particles, respectively (assuming in each
case that atomic nuclei of the other element are quantum particles).
We note that these results are similar to those presented earlier in
Ref. \onlinecite{ra08}, but here the simulations extended over
runs 10 times longer in order to improve the accuracy in the lattice 
parameter, and the present error bars are smaller by a factor of $\sim 3$. 

\begin{figure}
\vspace{-2.0cm}
\hspace{-0.5cm}
\includegraphics[width= 9cm]{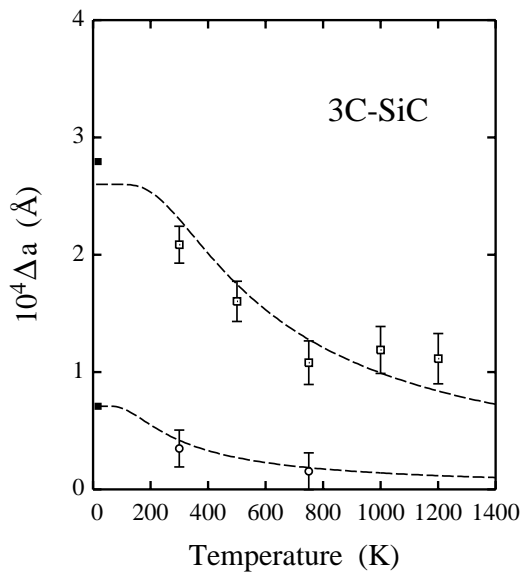}
\vspace{-2.5cm}
\caption{
Temperature dependence of the isotope effect on the lattice parameter
of 3C-SiC. Squares represent the difference between lattice parameter
of $^{\rm nat}$Si$^{12}$C and $^{\rm nat}$Si$^{13}$C, whereas circles
indicate the difference $\Delta a$ between $^{28}$Si$^{\rm nat}$C and
$^{29}$Si$^{\rm nat}$C. Filled symbols indicate the low-temperature values
derived from Eq.~(\ref{daa1}).
Lines represent fits of the simulation results to Eq.~(\ref{fit}).
}
\label{fig2}
\end{figure}

 Results presented until now were obtained for SiC crystals with natural
isotopic composition.
We now turn to simulations of SiC crystals, in which one of the elements
is taken to be isotopically pure. Thus, we carried out PIMD simulations
of $^{\rm nat}$Si$^{12}$C and $^{\rm nat}$Si$^{13}$C (where the superscript
``nat" refers to the mass of the natural isotopic composition) at several
temperatures. Results for the difference $\Delta a$ between lattice
parameters of both kinds of crystals are shown in Fig. 2 (squares).
At room temperature, we find $\Delta a = 2.1 \times 10^{-4}$ \AA,
which translates into a relative change $\Delta a/a = 4.8 \times 10^{-5}$.
Something similar has been done for $^{28}$Si$^{\rm nat}$C and
$^{29}$Si$^{\rm nat}$C crystals, and the difference $\Delta a$ 
yielded by our simulations is much lower: at 300 K we found 
$\Delta a = 3.5 \times 10^{-5}$ \AA, or 
$\Delta a/a = 8 \times 10^{-6}$.
The temperature dependence of $\Delta a$ can be fitted to a function 
with the shape of the mass derivative of a Bose-Einstein function, namely:
\begin{equation}
\Delta a = C \left[ 1 + \frac{2}{{\rm e}^x - 1} \left( 1 -
       \frac{x {\rm e}^x}{{\rm e}^x - 1} \right) \right] \,
\label{fit}
\end{equation}
where $x = T_a / T$, and $C$ and $T_a$ are fit parameters.
Dashed lines in Fig.~2 represent this function with 
$T_a$ = 1200 K (upper curve) and 600 K (lower curve).
Eq.~(\ref{fit}) is basically equivalent to Eq.~(3) of
Ref. \onlinecite{ra08} except that the adjustable parameter
$b$ has been replaced by well-defined physical variables more
convenient for the subsequent treatment. 
The minor deviations of the PIMD points from the fitted curves are
not surprising since we have used a single Einstein oscillator fit.
They could be decreased by using two oscillators.\cite{ma04}

The results of our PIMD simulations can be further analyzed in terms of 
a QHA for the lattice vibrations.  In such an approximation, the
lattice parameter $a(T)$ for a given isotopic composition at
temperature $T$ can be derived by minimizing the Helmholtz free energy 
with respect to the crystal volume.\cite{de96,ca05b}  One finds
\begin{equation}
a(T) = a_{\infty} + \frac{1}{ 3 B a^2_{\infty} }
   \sum_{n, {\bf q}}  \gamma_n({\bf q})  E_n({\bf q},T)
    \hspace{0.2cm}  ,
\label{amt}
\end{equation}
where
\begin{equation}
   E_n({\bf q},T) =  \frac{1}{2} \hbar \omega_n({\bf q})
   \coth \left( \frac{\hbar \omega_n({\bf q}) } {2 k_B T} \right)
    \hspace{0.2cm}   .
\label{ent}
\end{equation}
Here, $\omega_n({\bf q})$ are the frequencies of the $n$th
mode in the crystal, $B$ is the bulk modulus,
$a_{\infty}$ is the zero-temperature lattice parameter
 in the limit of infinite atomic mass (classical limit), and
$\gamma_n({\bf q}) = - \partial \ln \omega_n({\bf q}) /
 \partial \ln V $ is the Gr\"uneisen parameter of mode
$n, {\bf q}$.
Then, at $T = 0$ the difference $a(0) - a_{\infty}$ is given by
\begin{equation}
a(0) - a_{\infty} = \frac{1}{6 B a^2_{\infty}}
   \sum_{n, {\bf q}}  \hbar \omega_n({\bf q}) \gamma_n({\bf q})
    \hspace{0.2cm}   .
\label{aa1}
\end{equation}
Let us consider now for simplicity two isotopically-pure monatomic 
crystals with a mass difference $\Delta M$. The difference between
the corresponding lattice parameters, $\Delta a$, can be related
to the zero-point renormalization, $a_{\rm nat}(0) - a_{\infty}$, for
the natural crystal.  This can be achieved
through a first-order expansion for the lattice parameter as
a function of the mass $M$, and taking into account that the frequencies
$\omega_n({\bf q})$ scale as $1/\sqrt{M}$.
One finds for the change in lattice parameter at $T = 0$:
\begin{equation}
\Delta a = - \frac12 \left( a_{\rm nat}(0) - a_{\infty} \right) 
          \, \frac {\Delta M}{M_{\rm nat}}
        \hspace{0.2cm}   .
\label{daa1}
\end{equation}
This means that the low-temperature changes in $a$ due to isotopic mass
can be obtained from the zero-point renormalization of 
the lattice parameter in the natural crystal. 

For binary compounds such as SiC, one can use a formula similar to 
Eq.~(\ref{daa1}) to obtain the separate contributions of each kind
of atoms (i.e., Si or C).  To first order, the contributions of both
types of atoms will be  additive. Then, from the difference 
$a_{\rm nat}(0) - a_{\infty}$ discussed above, and presented in Fig. 1,
we obtain in the low-temperature limit, using Eq.~(\ref{daa1}), 
$\Delta a = 2.8 \times 10^{-4}$ \AA \, when replacing $^{12}$C by
$^{13}$C, and $\Delta a = 7.0 \times 10^{-5}$ \AA \, for substitution
of $^{28}$Si by $^{29}$Si. These are the values shown in Fig. 2 as
filled squares at T = 0.

\begin{figure}
\vspace{-2.0cm}
\hspace{-0.5cm}
\includegraphics[width= 9cm]{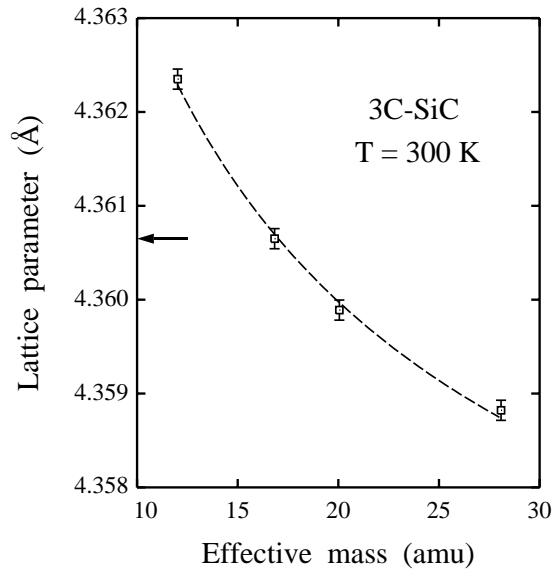}
\vspace{-2.5cm}
\caption{
Lattice parameter of 3C-SiC as a function of $M_{\rm eff}$ in a
virtual-crystal approximation, in which all atoms in the simulation cell
(Si and C) are assumed to have the same effective mass. Open squares
indicate results of PIMD simulations in this approximation. The dashed line
is a fit to Eq.~(\ref{abc}). An arrow indicates the lattice parameter
derived from PIMD simulations with C and Si atoms having their actual mean
isotopic masses.
}
\label{fig3}
\end{figure}

We note that the virtual-crystal approximation has been employed in our
simulations, i.e., for crystals with the natural isotopic composition
of Si or C we have assumed that the atoms have the average mass.
In this respect it is worthwhile considering the validity of assuming
an effective mass $M_{\rm eff}$ for all the atoms in a given crystal, to describe 
changes in the lattice parameter. 
In view of Eq.~(\ref{aa1}), one has:\cite{ca05b} 
\begin{equation}
\frac{a(0) - a_{\infty}}{a_{\infty}}  \sim \frac{\hbar}{B V_c}
      \langle \omega_n({\bf q}) \rangle  \langle \gamma_n({\bf q}) \rangle
    \hspace{0.2cm}   ,
\label{aa2}
\end{equation}
where $V_c$ is the volume of the primitive cell, and $\langle ... \rangle$
indicates an average over all branches of the Brillouin zone.
Now we may assume a dependence of the average frequency on effective mass as
\begin{equation}
  \langle \omega_n({\bf q}) \rangle \sim M_{\rm eff}^{-1/2}   ,
\end{equation}
which, in fact, is expected when one considers $M_{\rm eff}$ as 
\begin{equation}
   \frac{1}{M_{\rm eff}} = \frac12 \left( \frac{1}{M_{\text C}} +
            \frac{1}{M_{\text Si}}  \right) \, .
\label{meff}
\end{equation}
To check this point we have carried out PIMD simulations for SiC crystals
with various effective masses for both Si and C.
The results are displayed in Fig. 3 as open symbols. For comparison with the
result derived for $M_{\rm eff}$ given by Eq.~(\ref{meff}) ($\sim 16.8$ amu),
we also show those obtained by assuming an effective mass given either
by the average $(M_{\text C} + M_{\text Si})/2$, or by the mass of Si or C. 
All these results can be fitted well to the expression
\begin{equation}
  a = b + \frac{c}{\sqrt{M_{\rm eff}}}  \, ,
\label{abc}
\end{equation}
where $b$ and $c$ are fit parameters. Such an expression can be expected from
Eq.~(\ref{aa1}) when one considers an effective mass for both types of atoms,
at temperatures $T \ll \Theta_D$ (for SiC, $\Theta_D \sim 1100$ K).
The actual lattice parameter of SiC yielded by the simulations above is 
indicated in Fig. 3 by an arrow. It coincides within error bars with 
that derived assuming $M_{\rm eff} = 16.8$ amu, as given in Eq.~(\ref{meff}).
Note that taking the effective mass as the average of the masses (at about
20 amu) yields a lattice parameter clearly lower than the actual one
obtained using the separate masses of C and Si.

The agreement between the result for $M_{\rm eff} = 16.8$ amu and the
real crystal can be interpreted in terms of perturbation theory as follows.
Looking at Eq.~(\ref{aa1}), changes in the lattice parameter are mainly 
due to TO phonons, as can be derived from Fig.~9 in Ref.~\onlinecite{ka94}
for the appropriate values of $\gamma_n$ and Fig.~2 for the density of
phonon states.
The TO band in SiC is rather symmetric,\cite{ka94} and according to 
second-order perturbation theory,\cite{wi02,ca05b} mass fluctuations cause 
an increase in the high frequencies 
and a reduction of the low ones by a similar amount. All together the
effect in the lattice parameter is expected to be negligible, as observed 
in the results of the simulations.
If some effect appears due to mass fluctuations, it has to be of third or
higher order, and should be less than the statistical uncertainty of
our results.
This gives further support to the virtual-crystal approximation for
calculating lattice parameters of this kind of semiconductors.

In summary, we have calculated the isotope effect on the lattice parameter
of cubic SiC by PIMD simulations. This procedure gives a quantitative 
estimation of such effect, which amounts to $\Delta a/a = 4.8 \times 10^{-5}$
for replacement of $^{12}$C by $^{13}$C, and to 
$\Delta a/a = 8 \times 10^{-6}$ for substitution of $^{29}$Si for
$^{28}$Si. These results have been interpreted in terms of a quasi-harmonic
approach for the lattice vibrations.
The virtual-crystal approximation is found to be valid in all cases
considered here.
Second order perturbation theory, as implied by Eq.~(\ref{daa1}), has also 
been shown to be valid.

\begin{acknowledgments}
This work was supported by Ministerio de Ciencia e Innovaci\'on (Spain)
through Grant No. FIS2006-12117-C04-03 and by CAM through project 
S-0505/ESP/000237.
\end{acknowledgments}

\end{document}